# Impacts of Atomistic Coating on Thermal Conductivity of Germanium Nanowires

Jie Chen,[1] Gang Zhang,[2,*] and Baowen Li[1,3]

[1]Department of Physics, Centre for Computational Science and Engineering, and Graphene Research Centre, National University of Singapore, Singapore 117542, Singapore

[2] Key Laboratory for the Physics and Chemistry of Nanodevices and Department of Electronics, Peking University, Beijing 100871, People's Republic of China

[3]NUS-Tongji Center for Phononics and Thermal Energy Science, Department of Physics, Tongji University, Shanghai 200092, People's Republic of China

## Abstract

By using non-equilibrium molecular dynamics simulations, we demonstrated that thermal conductivity of Germanium nanowires can be reduced more than 25% at room temperature by atomistic coating. There is a critical coating thickness beyond which thermal conductivity of the coated nanowire is larger than that of the host nanowire. The diameter dependent critical coating thickness and minimum thermal conductivity are explored. Moreover, we found that interface roughness can induce further reduction of thermal conductivity in coated nanowires. From the vibrational eigen-mode analysis, it is found that coating induces localization for low frequency phonons, while interface roughness localizes the high frequency phonons. Our results provide an available approach to tune thermal conductivity of nanowires by atomic layer coating.

*zhanggang@pku.edu.cn

Keywords: Thermal conductivity; Germanium nanowires; Atomistic coating; Molecular dynamics simulations

A novel nanostructure – core-shell nanowires (NWs) recently comes to the spotlight of research. Compared to the single component NWs, the physical properties of core-shell NWs can be further tailored by different compositions and different configurations, thus leading to various novel nanoscale electronic and optoelectronic devices [1-3]. In addition to the electronic and optical properties, thermal property of NWs has attracted more and more interests due to the potential thermoelectric applications in both power generation and refrigeration [4]. For instance, by etching the surface of silicon NWs, it has been experimentally demonstrated that thermal conductivity of Si NW can be reduced more than two orders of magnitude compared with bulk Silicon [5, 6]. Moreover, remarkable reduction of thermal conductivity in core-shell [7-10], tubular [11, 12], and surface-decorated [13] NWs has also been reported through various kinds of interface and surface engineering.

Previous studies [7-9] on the reduction of thermal conductivity in core-shell NWs mainly focus on Si/Ge core-shell NWs, which is quite intuitive as Ge is a low thermal conductivity material compared to Si [14]. However, in experimental realizations, the NWs synthesised are Ge/Si core-shell NWs [1-3]. Very recently, both theoretical and experimental works have shown that thermal conductivity of Ge/Si core-shell NWs can be lower than that of pure GeNWs with the same cross section area [15, 16]. The reduction is caused by the localization of the longitudinal phonon modes induced by the coherent resonance effect between the transverse and longitudinal modes [15]. Thus the core-shell structure offers the unique opportunity to further reduce thermal conductivity of low thermal conductivity material even by coating with high thermal conductivity material.

Despite these recent progresses about the thermal properties of core-shell NWs, many important and fundamental issues remain unsolved. For example, in our recent work [15], thermal conductivity of Ge/Si core-shell NW is compared with that of pure GeNW with the same entire cross section area. However, thermal conductivity of NW is sensitive to the surface-to-volume ratio (SVR) and increases with the cross section area [17, 18]. Coating another material outside the host NW increases the cross section area and thus can lead to the increase of thermal conductivity. Therefore, to study the effect of coating on thermal conductivity from the experimental point of view, one should compare thermal conductivity of GeNWs before and after coating, i.e., thermal conductivity of the pristine GeNWs and resultant Ge/Si core-shell NWs.

Moreover, it remains unclear how does the reduction of thermal conductivity by coating depend on the host nanowire diameter, coating thickness, and the interface quality, which are important factors in real experiment to tailor the thermal conductivity of nanowires. In this letter, we address these questions by systematically investigating thermal conductivity of Si-coated Ge NWs by using molecular dynamics simulations, and comparing the results with that of pristine GeNWs before coating. Our focus here is to provide a direct design guidance for experiments.

The configuration of (100) GeNW is shown in Fig. 1a. The cross section of GeNW is square, with side length (diameter) denoted by $D_{Ge}$. The longitudinal direction is set along $x$ axis, with $Nx$, $Ny$, and $Nz$ unit cells (8 atoms per unit cell) in $x$, $y$, and $z$ direction, respectively. GeNW is uniformly coated with Si layer and the coating thickness is denoted by $D_{coating}$. The largest GeNW considered in our study has the diameter of $D_{Ge}$=11.3 nm, and the largest coating thickness is $D_{coating}$=3.8 nm. The largest simulation domain in our study has a total number of 147968 atoms. Stillinger–Weber (SW) potential [19, 20] is used to derive the force term, which has been widely used in the study of heat conduction in low-dimensional materials [21]. The force field parameters for Si-Si bond and Ge-Ge bond are from Ref. [19] and Ref.

[20], respectively. For Si-Ge bond, the net length and energy units in SW potential are taken to be the arithmetic average and geometric average of that of Si and Ge atom, respectively.

Because of the oscillation in heat current autocorrelation function (HCACF) observed in core-shell nanostructures [15], it is challenging to specify the thermal conductivity by applying the Green-Kubo method. Thus, we use non-equilibrium molecular dynamics (NEMD) simulations to calculate the thermal conductivity. In NEMD simulations, the velocity Verlet algorithm is employed to integrate Newton's equations of motion numerically. The canonical ensemble molecular dynamics with Langevin heat reservoir [22] first runs for $10^5$ steps with a time step of 0.8 fs to equilibrate the whole system at a given temperature, during which the free boundary condition is applied to the surface atoms in all directions. The lattice constant is then calculated according to the relaxed structure.

After structure relaxation, at the two ends of the NWs in the longitudinal direction, fixed boundary condition is imposed on the boundary layers. Next to the boundary layers at two ends, multiple layers of the NWs are put in contact with Langevin heat reservoir. The temperature of two heat reservoirs are set as $T_H=T_0+\Delta/2$ and $T_L=T_0-\Delta/2$, respectively, where $T_0$ is the mean temperature and $\Delta$ is the temperature difference. Free boundary condition is applied to atoms on the outer surface of the NWs. The simulations are then performed long enough ($2\times 10^7$ time steps) to allow the system to reach the non-equilibrium steady state where the temperature gradient is well established and the heat current going through the system is time independent. Thermal conductivity is calculated as:

$$\kappa = -J_L/\nabla T, \quad (1)$$

where $\nabla T$ is the temperature gradient, and $J_L$ is the heat current along the longitudinal direction defined as the energy transported along the NW in unit time through the unit cross section area. The temperature gradient is calculated according to the slope of the linear fit line of the local temperature at each layer and the lattice constant calculated after structure relaxation, which takes into account the variation of lattice constant for different core-shell structures.

Recent studies have shown that thermal conductivity in core-shell NWs is not sensitive to temperature [10, 23]. In this work, we mainly focus on the effect of diameter and coating thickness on thermal conductivity. Therefore, we fix the simulation temperature at room temperature in our study, and do not consider quantum correction [24]. Furthermore, the length dependent thermal conductivity in nanostructures has been reported by numerous studies [25, 26]. In this paper, our focus is the relative reduction of thermal conductivity by atomistic coating, rather than the absolute value or the length effect on thermal conductivity. Therefore, we fix the length of the nanowire (16 unit cells) in our simulations and mainly consider the variation of cross section area of GeNW and coating thickness.

To study the reduction of thermal conductivity before and after coating for Ge NWs with varying diameters, we increase the cross sectional side length (diameter) $D_{Ge}$ from 2.8 nm to 11.3 nm. In Fig. 1b, we show the room temperature thermal conductivity $\kappa$ of GeNWs ($D_{coating}$=0) and Ge/Si core-shell NWs ($D_{coating}$>0) for different $D_{Ge}$. It is obvious that for each GeNW, the coating of Si atomistic layers can reduce the thermal conductivity. Further coating leads to an increase of thermal conductivity. When the coating thickness is greater than a certain critical value, thermal conductivity of the resultant Ge/Si core-shell NWs becomes larger than that of pristine GeNW without coating.

Increasing coating thickness has two opposite effects on thermal conductivity. On the one hand, the creation of core-shell structures will induce the phonon resonance between the transverse and longitudinal modes, thus offering a coherent mechanism to reduce thermal conductivity. To provide detailed evidence of phonon resonance, we record HCACF along the longitudinal direction in Ge nanowires before and after coating by using equilibrium molecular dynamics (EMD) simulations. After structure relaxation, microcanonical ensemble EMD runs for $3\times10^6$ time steps to calculate HCACF. More details about the calculations of HCACF can be found in Ref. [15]. Fig. 2 shows the normalized HCACF for GeNWs with $D_{Ge}$=5.64 nm and Ge/Si core-shell NWs with $D_{coating}$=0.54 nm. HCACF in pristine GeNWs decays to zero after the relaxation time, and fluctuates around zero due to the computational noise. In contrast, HCACF in Si-coated GeNWs oscillates periodically around zero after the relaxation time (Fig. 2b), and this oscillation does not vanish for a very long time (Fig. 2a). The fast Fourier transform (FFT) of the oscillation signal shows that there are multiple resonant peaks in Si-coated GeNWs, with the lowest resonance frequency located at 0.53±0.04 THz (Fig. 2c). In contrast, the FFT amplitude in pristine GeNWs is two orders of magnitude smaller than that in Si-coated GeNWs, and no distinct resonant peak can be observed in the entire frequency regime, showing the characteristic of the noise spectrum. Moreover, it has been demonstrated in Ref. 15 that this lowest resonance frequency corresponds to the lowest eigen-frequency of the transverse modes in such nanowire structure calculated from the elastic medium theory.

In core-shell nanowires, atoms on the same cross section plane have different sound velocity in the longitudinal direction. As a result, atoms near the interface are stretched, which induces a strong coupling between the transverse and longitudinal motions. Due to the spatial confinement on the cross section plane, the transverse phonons are quantized and non-propagating, similar to the standing wave characteristic of the transverse modes in the acoustic waveguide [27]. The coupling between the transverse and longitudinal motions results in the resonance when the frequency of the longitudinal phonon mode is close to the eigen-frequency of the transverse mode [15]. Moreover, as the transverse phonons are non-propagating, this resonance effect results in longitudinal phonon localization, thus offering a coherent mechanism to reduce thermal conductivity.

On the other hand, the coating layers increase the entire cross section area. Due to the high SVR in nanowires compared to the bulk materials, thermal conductivity of nanowires depends on the cross section area as a consequence of the surface scattering of phonons [17, 18]. Due to the increase of the entire cross section area after coating, the SVR decreases and thus the strength of the surface scattering is reduced [18], leading to the increase of thermal conductivity with the increase of the coating thickness. Therefore, thermal conductivity of Si-coated GeNWs is determined by these two competing effects. When the coating thickness is less than a certain critical value, the suppression of the longitudinal phonon transport is the dominating factor, corresponding to the reduction of thermal conductivity. However, when the coating layers increase further beyond the critical thickness, the reduced surface scattering dominates the thermal transport, leading to the enhanced thermal conductivity compared to that of pristine GeNW.

Our results show that in practical application, one can control thermal conductivity of NWs by coating other atoms. This approach offers novel avenues and more flexibility for the design and thermal management in nanostructures. For instance, small diameter NWs are favourable for thermoelectric applications due to the low thermal conductivity, but are more challenging for experimental synthesis. Through the coating method, the low thermal conductivity feature close to the very thin NW can be obtained from a much thicker NW. This point can be seen from Fig. 1b. Thermal conductivity of pristine GeNWs with $D_{Ge}$=9.0 nm is close to that of coated GeNWs with $D_{Ge}$=11.3 nm and $D_{coating}$=2.7 nm. The resultant core-shell NW has a cross sectional side length ($D_{Ge}+2*D_{coating}$) of 16.7 nm, which is almost twice of the cross sectional side length for pristine GeNW with the similar thermal conductivity. Moreover, for NWs with a given diameter, the coating method can preserve the low thermal conductivity feature of the original NWs up to certain critical coating thickness. For example, for NW with $D_{Ge}$=9.0 nm, coating layers less than 2.7 nm can give rise to thermal conductivity lower than that of pristine GeNW. As this coating thickness is achievable with atomistic layer deposition technology, our results indicate a practical approach that can be realized to tune thermal conductivity.

As shown in Fig. 1b, the critical coating thickness depends on the diameter of the NW before coating. To see this dependence more clearly, we plot in Fig. 1c the normalized thermal conductivity versus coating thickness for GeNWs with different diameters. Thermal conductivity of GeNW ($D_{coating}$=0) is used as reference at each $D_{Ge}$. According to the above-mentioned two competing effects, the maximum reduction of thermal conductivity occurs at the optimal coating thickness of about 0.5~1 nm (1~2 unit cell). Further increase of coating thickness will result in the increase of thermal conductivity, but the increase trend becomes slower at larger diameters. These findings are consistent with a recent study on thermal conductivity reduction in Si/Ge core-shell NWs [9]. The dashed line in Fig. 1c draws the unity for reference, and the arrows point the critical coating thickness below which coating is effective for the purpose of thermal conductivity reduction. Figure 1d shows the dependence of critical coating thickness $D_{critical}$ on the NW diameter: it monotonically increases with NW diameter, with a linear slope of about 0.34. These results suggest that coating is a quite effective method for thermoelectric applications and is more robust at larger diameter.

Figure 3 shows the minimum thermal conductivity of Ge/Si core-shell NWs versus the cross sectional side length of GeNW. Thermal conductivity of the uncoated GeNW is used as reference. When the NW diameter increases, the impact of coating on the reduction of thermal conductivity increases, and the normalized thermal conductivity converges to about 74% at large diameter (see caption of Fig. 3), corresponding to a 26% reduction. Note this reduction ratio is smaller than previous reported value [15] because thermal conductivity of the GeNW before coating is now used as reference, rather than that of GeNW with the same entire cross sectional area after coating. We should point out that with the increase of NW diameter, although the maximum reduction ratio of thermal conductivity by coating is higher, thermal conductivity of host NW also increases with diameter. Therefore, thin NW is still more favourable for thermoelectric application, because the absolute value of the minimum thermal conductivity of the coated NW is increased with diameter (see Fig. 1b).

So far, we have only considered the impact of atomistic coating with perfect core-shell interface. In practical conditions, both the NW and coating process are impossibly smooth in atomistic scale. Thus in the following, we will study the effect induced by imperfect interface, i.e., interface roughness. In our study, we introduce the interface roughness by randomly switching certain percentage of the atoms at the core-shell interface. Fig. 4a shows the cross sectional view of (100) Ge/Si core-shell NWs with different interfacial quality. The green and yellow circles denote Ge and Si atoms, respectively. The blue box draws the boundary for the perfect interface. For perfect interface, all the Ge atoms are inside the blue box and all the Si atoms are outside the box, giving rise to a sharp interface. After random-switch, some of the Ge atoms are outside the box and some Si atoms are inside the box, resulting in the rough interface. For this type of roughness, we only consider a very small fluctuation of about 1~2 atomic layer in thickness (less than 3 Å) at the interface.

To check the generality of the phonon resonance effect in core-shell nanowires, Fig. 2 also shows the normalized HCACF for Ge/Si core-shell NWs with rough interface. The periodic oscillation feature in HCACF remains intact, with insignificant difference in HCACF among different interfacial quality. Furthermore, multiple resonant peaks also exist in the FFT of the oscillation signal for the core-shell NWs with rough interface, and the resonance frequency remains the same, especially for the low-frequency resonant peaks (Fig. 2c). These results suggest that the phonon resonance is a generic feature in core-shell nanowires and is robust to interface quality.

Fig. 4b shows thermal conductivity of Ge/Si core-shell NWs with different interfacial quality versus coating thickness. For each random-switch percentage, six nanowire samples with different interface structures are calculated in our simulations. The empty square draws thermal conductivity of GeNWs without coating ($D_{Ge}$=5.6 nm) for comparison. The interface roughness does not change the qualitative dependence of thermal conductivity on coating thickness, but further reduces thermal conductivity of Ge/Si core-shell NW at a given coating thickness. Moreover, this additional reduction of thermal conductivity increases monotonically with the interface roughness (random-switch percentage). The maximum reduction ratio for the GeNW with $D_{Ge}$=5.6 nm increases from 17% for perfect interface to 42% for rough interface with 50% atoms randomly switched (see Fig. 4c). This monotonic decreasing result is consistent with previous theoretical works [28, 29] about the impact of surface roughness on thermal conductivity of thin NWs. Therefore, with the introduction of interface roughness, the critical coating thickness $D_{critical}$ can be further extended (Fig. 4c). The recent experimental study by Kang *et al.* [30] has found that the interface roughness of core-shell nanowires can be engineered in experiment by applying different deposition power, with high power resulting in the rough interface. Moreover, their experimental study has reported the reduction of thermal conductivity in Bi/Te core-shell nanowires with rough interface, which is consistent with our results that interface roughness leads to further reduction of thermal conductivity in core-shell nanowires.

To understand the mechanism of thermal conductivity reduction in coated NWs with different interface quality, we carry out a vibrational eigen-mode analysis on GeNWs before and after coating. Mode localization can be quantitatively characterized by the participation ratio $P_\lambda$ defined for each eigen-mode $\lambda$ as [31]

$$P_\lambda^{-1} = N \sum_i \left( \sum_\alpha \varepsilon_{i\alpha,\lambda}^* \varepsilon_{i\alpha,\lambda} \right)^2, \qquad (2)$$

where $N$ is the total number of atoms, and $\varepsilon_{i\alpha,\lambda}$ is the $\alpha$-th eigenvector component of eigen-mode $\lambda$ for the $i$-th atom. The participation ratio (p-ratio) measures the fraction of atoms participating in a given mode, and effectively indicates the localized modes with $O(1/N)$ and delocalized modes with $O(1)$. It can provide detailed information about localization effect for each mode.

Figure 5 compares the p-ratio for each eigen-mode in pristine GeNWs (black), Ge/Si core-shell NWs with perfect interface (red), and Ge/Si core-shell NWs with different percentage of interface roughness (blue and green). All phonon modes are computed by using the "General Utility Lattice Program" (GULP) package [32] with the cross sectional side length $D_{Ge}$=5.64 nm and coating thickness $D_{coating}$=0.54 nm.

After Si-coating, additional phonon modes (>10 THz) are present in the high frequency regime (Fig. 5a) due to the extended frequency spectrum of Si compared to Ge [33]. In the low frequency regime, there is a reduction of p-ratio in coated NW compared to pristine GeNW. This phonon localization effect is caused by the resonance effect induced by coupling between the transverse and longitudinal phonon modes in core-shell nanostructures [15]. As the strongest resonance peak is related to the lowest eigen-frequency in the transverse direction, the localization is remarkable for phonons with long wavelength (larger than the NW diameter).

As Si and Ge have different cut-off frequency for acoustic phonons [33], the Si-coating induces the extension of high frequency acoustic phonons in Si-coated GeNWs. As a result, the participation ratio in Si-coated GeNWs appears to be higher than that in pristine GeNWs for those phonons with frequency corresponding to the Brillouin zone edge of GeNWs. Since the phonon resonance is a coherent process, its effect on phonon localization is more significant for low frequency acoustic phonons near the Brillouin zone center. To demonstrate the effect of phonon resonance on the mode localization more clearly, the polarization-resolved participation ratio for the longitudinal acoustic (LA) phonons near the Brillouin zone center is shown in the inset of Fig. 5a. The participation ratio in pristine GeNWs decreases monotonically with the increase of frequency, due to non-propagating nature of high frequency phonons near the Brillouin zone edge. After coating, the participation ratio in Si-coated GeNWs shows an overall reduction for very low frequency LA phonons near the Brillouin zone center. More importantly, an obvious dip of the participation ratio is observed at around 0.5 THz in Si-coated GeNWs, which is consistent with the resonance frequency found in FFT of HCACF along the longitudinal direction (Fig. 2c) and thus provides the strong evidence that phonon resonance in Si-coated GeNWs leads to localization for low frequency phonons near the Brillouin zone center.

After the interface roughness is introduced, the p-ratio of the low-frequency phonons less than 4 THz, corresponding to the acoustic phonons in bulk crystals [33, 34], is almost unaffected, and the resonance induced phonon localization effect in this regime remains intact, which is consistent with the HCACF shown in Fig. 2. More interestingly, the additional reduction of p-ratio in the high frequency regime (>10 THz) is observed in Ge/Si core-shell NWs with rough interface (Fig. 5b and 5c). The impact of rough edge on thermal conductivity of NWs [28, 29, 35-37] and graphene nanoribbons [38] has been widely studied. It is found that the modes with the wavelength less than the NW diameter are localized by the rough edges. This is consistent with our vibrational eigen-mode analysis that interface roughness mainly induces the reduction of the p-ratio in high frequency regime. Phonons with frequency higher than 12 THz correspond to the optical phonons in (100) Si [34]. For bulk crystal, due to the low group velocity and short phonon lifetime of optical phonons, their contributions to thermal conductivity are very small [39, 40], so that they are usually neglected. However, recent study based on first-principles calculations reported optical phonons can contribute over 20% to the thermal conductivity of nanostructures at room temperature [41]. The interface roughness results in the additional localization effect of optical phonons, thus leading to the further reduction of thermal conductivity compared to the perfect interface (Fig. 4).

The results in Figure 5 also help to understand the dramatically low thermal conductivity observed experimentally in rough SiNWs [5]. To explain this low thermal conductivity, different phenomenological models were suggested in literature [28, 29, 35-37]. For instance, a NW is described as a system composed of a finite number of interacting one-dimensional chains with the longitudinal oscillations, in which edge roughness is modeled by randomly missed atoms in the edge chains [36]. Moreover, phonon backscattering is simulated by Monte Carlo simulation for sawtooth roughness on a NW [28]. Although the low thermal conductivity of rough SiNW has been qualitatively explained by the surface scattering mechanism, there exists a discrepancy between the NW used in experiment and the theoretically modeled NW. The etched SiNWs used in experiment are in fact coated with the amorphous $SiO_2$ layers [5], i.e., they are essentially core-shell NWs. However, all the NW models in these theoretical studies are pristine SiNWs without any coating layer [28, 29, 35-37]. From our results, it is obvious that both coating and interface roughness can induce the reduction of thermal conductivity. More importantly, their impacts on phonon spectrum are separated. Coating induces localization for low frequency phonons (wavelength comparable to NW diameter), while interface roughness localizes the high frequency phonons (wavelength is much less than NW diameter). So including the coating effect can cause further reduction of thermal conductivity. Moreover, as the coating layers in Ref. 5 are amorphous $SiO_2$, amorphous layer can induce significant decrease of both phonon group velocity and lifetime compared to the crystalline NW [42]. This suggests that the combined effects of the surface coating, interface roughness and amorphous layers are responsible for the ultralow thermal conductivity observed experimentally in rough SiNWs [5]. Ignoring the effects of coating and amorphous layers may be the partial origin for the discrepancy between experimental measurement and numerical predictions.

In summary, we have demonstrated that Ge NWs with surface atomistic coating are promising low thermal conductivity materials. There is a critical coating thickness beyond which thermal conductivity of the coated NW is larger than that of the host NW. When the NW diameter increases, the impact of coating on the reduction of thermal conductivity increases, and the normalized room temperature thermal conductivity converges to about 74% at large diameter. Furthermore, we have also studied the effect induced by imperfect interface, i.e., interface roughness. Our results show that the interface roughness further reduces thermal conductivity of coated NW at a given coating thickness. From the vibrational eigen-mode analysis, it is obvious that surface coating induces localization for phonons with wavelength comparable to NW diameter, while interface roughness localizes the phonons with wavelength much less than NW diameter. So including the coating effect can cause further reduction of thermal conductivity in nanostructures. Combined with the availability of atomistic layer deposition technology, our results indicate a practical approach that can be realized experimentally to tune thermal conductivity.

## Acknowledgements

The work has been supported by a grant W-144-000-280-305 from SERC of A*STAR, Singapore. GZ was supported by the Ministry of Science and Technology of China (Grant Nos. 2011CB933001).

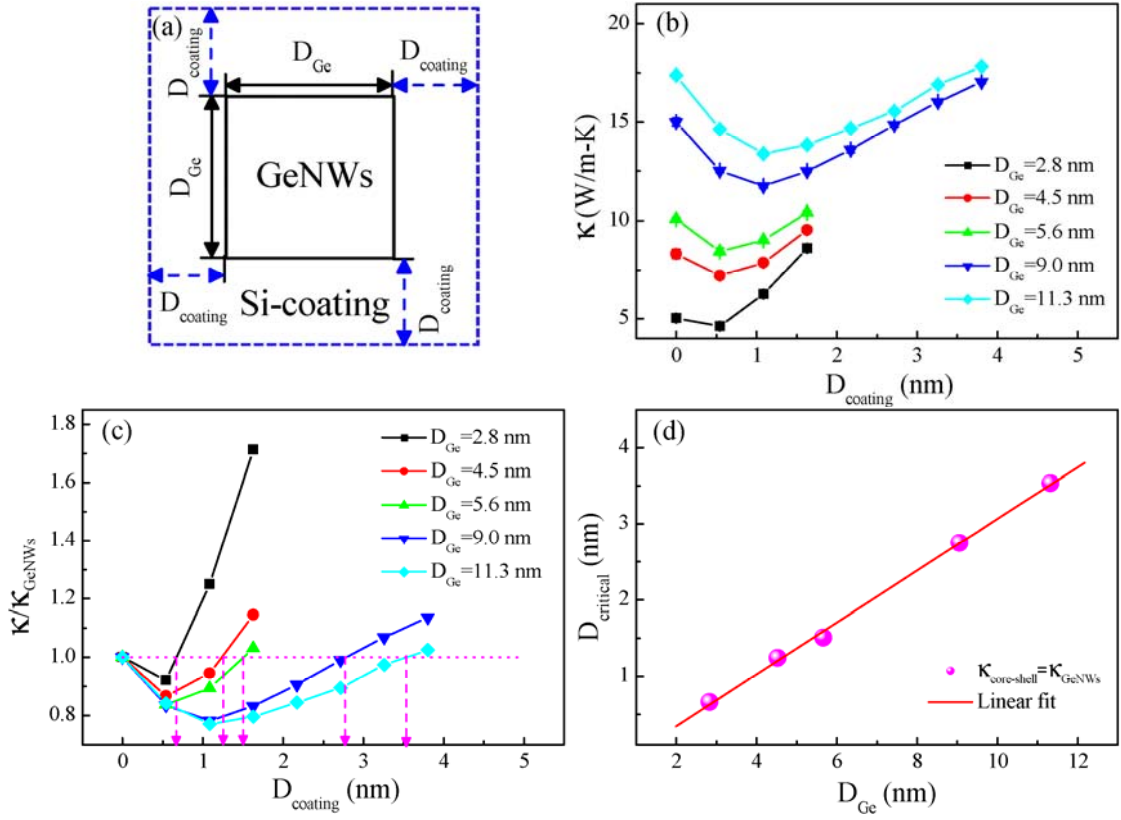

Figure 1. Coating configuration and thermal conductivity. (a) Cross section of (100) GeNWs (solid line) with Si-coating shell (dashed line). $D_{Ge}$ denotes the cross sectional side length of GeNWs, and $D_{coating}$ denotes the thickness of Si-coating. (b) Thermal conductivity of GeNWs ($D_{coating}=0$) and Ge/Si core-shell NWs ($D_{coating}>0$) for different $D_{Ge}$ at room temperature. The nanowire length is fixed as 16 unit cells. (c) Normalized thermal conductivity versus coating thickness for different $D_{Ge}$. Thermal conductivity of GeNWs at each $D_{Ge}$ is used as reference. The dashed arrows point the critical coating thickness when thermal conductivity of Ge/Si core-shell NWs ($\kappa_{core-shell}$) is equal to that of GeNWs ($\kappa_{GeNWs}$). The dashed line is drawn to guide the eye. (d) The critical coating thickness $D_{critical}$ ($\kappa_{core-shell}=\kappa_{GeNWs}$) versus cross sectional side length of GeNWs. The solid line draws the linear fit line.

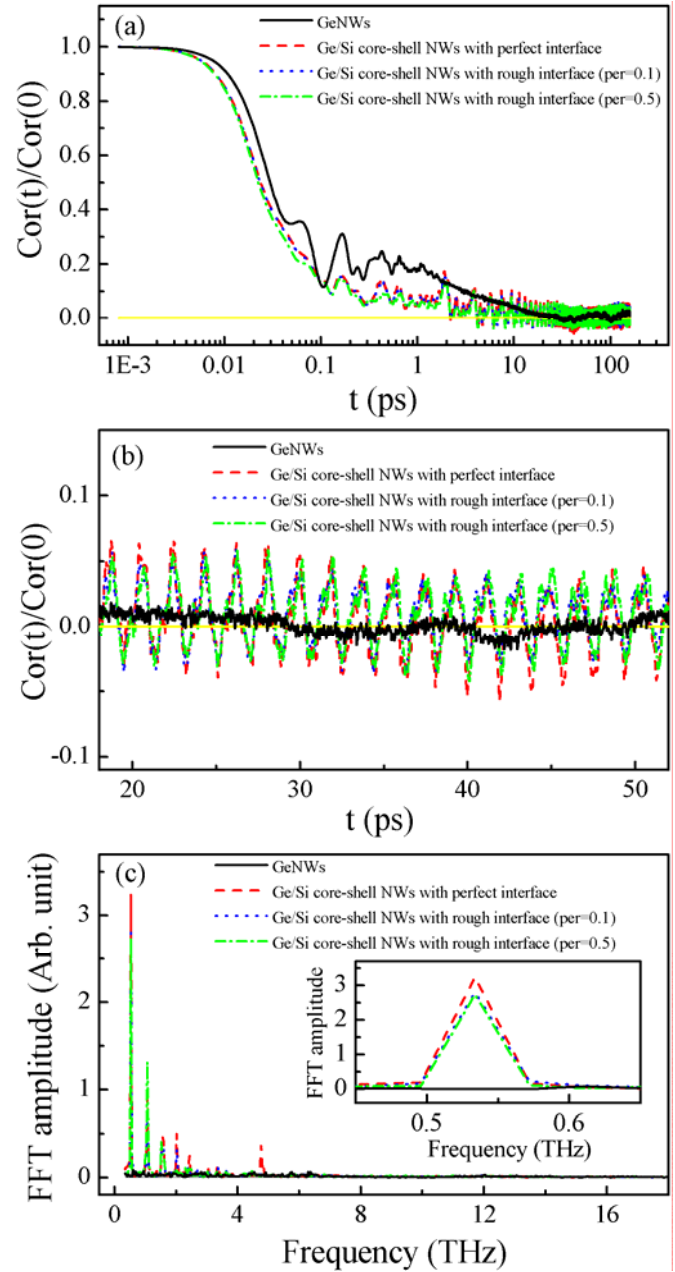

Figure 2. The normalized heat current autocorrelation function and its fast Fourier transform (FFT). (a) Normalized heat current auto-correlation function for GeNWs and Ge/Si core-shell NWs with different interface quality. The solid yellow line draws zero axis for reference. (b) Long-time region of (a). (c) FFT of the normalized correlation function shown in (b). The inset in (c) zooms in for the resonance peak with the lowest frequency. Here we show calculation results for (100) GeNWs with $D_{Ge}$=5.64 nm and Ge/Si core-shell NWs with $D_{coating}$=0.54 nm.

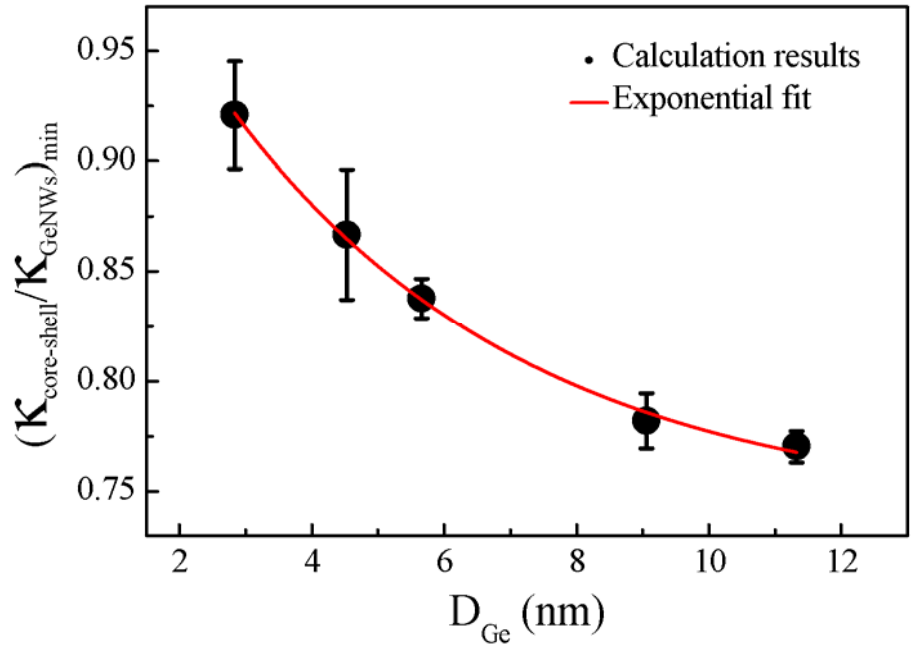

Figure 3. Minimum of the normalized thermal conductivity of Ge/Si core-shell NWs versus cross sectional side length $D_{Ge}$ of GeNWs. Thermal conductivity of GeNWs at each $D_{Ge}$ is used as reference. The solid line draws the exponential fit curve according to $min=A_0*exp(B_0*D_{Ge})+C_0$, yielding $A_0=0.34\pm0.02$, $B_0=-0.22\pm0.03$, and $C_0=0.74\pm0.01$.

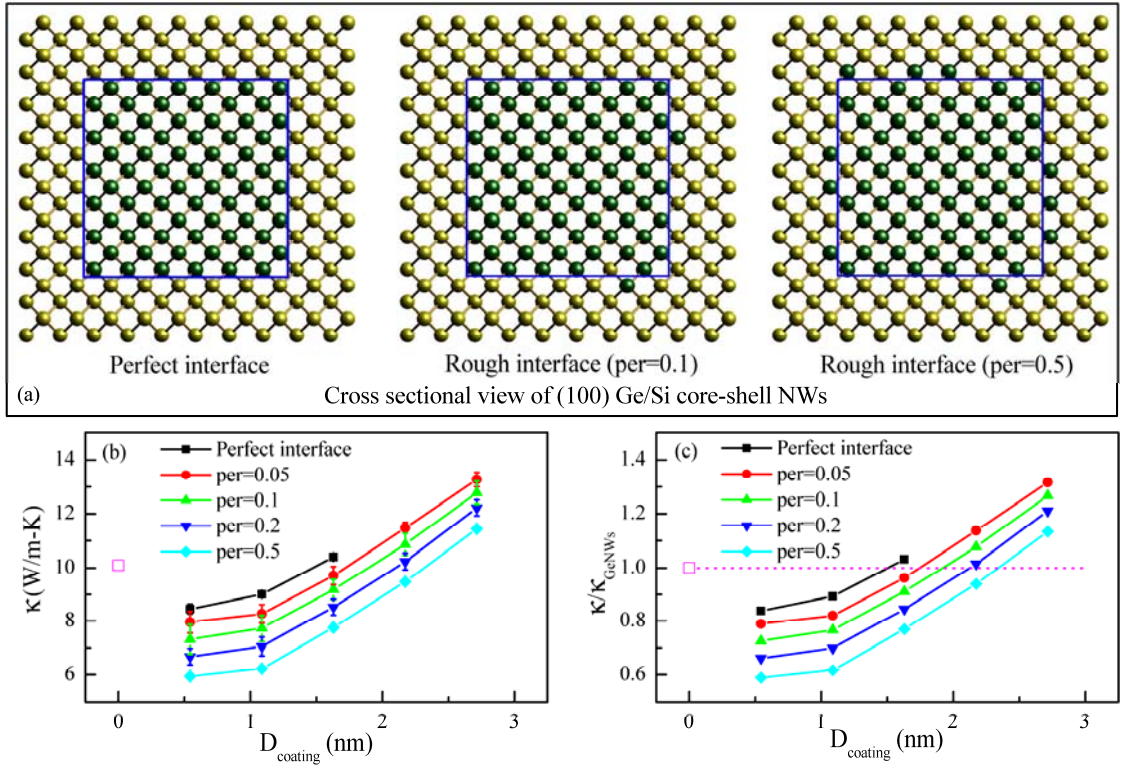

Figure 4. Core-shell interfacial quality and thermal conductivity. (a) Cross sectional view of (100) Ge/Si core-shell NWs. The green and yellow circles denote Ge and Si atoms, respectively. The left, central and right panel shows, respectively, the perfect interface, the rough interface with 10% of interfacial atoms (per=0.1) randomly switched, and the rough interface with 50% of interfacial atoms (per=0.5) randomly switched. The blue box draws the boundary for the perfect interface. (b) Thermal conductivity of Ge/Si core-shell NWs versus coating thickness for different interfacial quality. For each random-switch percentage, six nanowire samples with different interface structures are used in simulations. The empty square shows thermal conductivity of GeNWs before coating, which has a cross sectional side length of 5.6 nm and a longitudinal length of 9.0 nm. (c) Normalized thermal conductivity versus coating thickness for different interfacial quality. Thermal conductivity of GeNWs is used as reference. The dashed line is drawn to guide the eye.

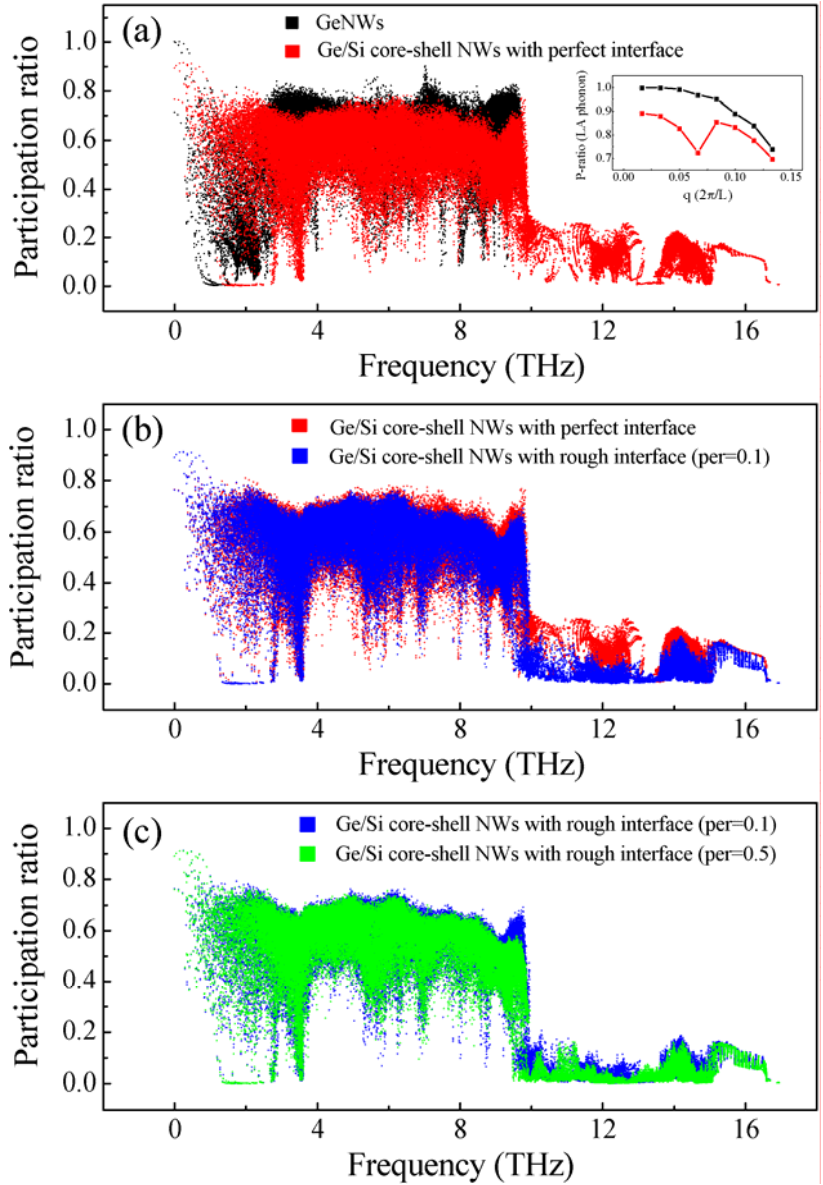

Figure 5. Participation ratio for different phonon modes in GeNWs before and after coating. The black, red, blue and green colors denote, respectively, participation ratio in GeNWs, Ge/Si core-shell NWs with perfect interface, Ge/Si core-shell NWs with 10% interfacial roughness, and Ge/Si core-shell NWs with 50% interfacial roughness. Here we show calculation results for (100) GeNWs with $D_{Ge}$=5.64 nm and Ge/Si core-shell NWs with $D_{coating}$=0.54 nm. The inset in (a) shows the polarization-resolved participation ratio (p-ratio) for the longitudinal acoustic (LA) phonon near the Brillouin zone center in GeNWs (black) and Ge/Si core-shell NWs with perfect interface (red).